\documentclass[12pt]{article}

\usepackage{graphicx}

\title{ Exact relativistic Green's functions for the time-independent potentials }
\author{  Yu.A.Simonov \\
State Research
Center\\Institute of Theoretical and Experimental Physics, \\
Moscow, 117218 Russia}

\newcommand{\beq}{\begin{eqnarray}}
 \newcommand{\eeq}{\end{eqnarray}}
\newcommand{\be}{\begin{equation}}
 \newcommand{\ee}{\end{equation}}

 \def\la{\mathrel{\mathpalette\fun <}}

\def\fun#1#2{\lower3.6pt\vbox{\baselineskip0pt\lineskip.9pt
\ialign{$\mathsurround=0pt#1\hfil ##\hfil$\crcr#2\crcr\sim\crcr}}}

\newcommand{{\SD}}{\rm SD}

\newcommand{{\Mc}}{\mathcal{M}}

\newcommand{\vex}{\mbox{\boldmath${\rm x}$}}
\newcommand{\vey}{\mbox{\boldmath${\rm y}$}}
\newcommand{\ver}{\mbox{\boldmath${\rm r}$}}
\newcommand{\vesig}{\mbox{\boldmath${\rm \sigma}$}}

\newcommand{\veP}{\mbox{\boldmath${\rm P}$}}
\newcommand{\vep}{\mbox{\boldmath${\rm p}$}}

\newcommand{\vez}{\mbox{\boldmath${\rm z}$}}

 \newcommand{\veA}{\mbox{\boldmath${\rm A}$}}

\newcommand{\ves}{\mbox{\boldmath${\rm s}$}}
\newcommand{\vek}{\mbox{\boldmath${\rm k}$}}

\newcommand{\veB}{\mbox{\boldmath${\rm B}$}}

\newcommand{\veE}{\mbox{\boldmath${\rm E}$}}

\newcommand{\veal}{\mbox{\boldmath${\rm \alpha}$}}

\newcommand{\lan}{\langle}
\newcommand{\ran}{\rangle}

\begin{document}
\maketitle
\begin{abstract}
  Relativistic formalism of Green's functions   is discussed in QCD and QED,  where the relativistic Green's functions are constructed using the Schwinger proper time formalism and  analysed using the Fock--Feynman--Schwinger method. As a result a simple and exact method  is found for the relativistic systems, where the  interaction can be written in a  time-independent form.   In this case one can write the relativistic Green's function   as  a one-dimensional integral of the corresponding nonrelativistic Green's function.  The explicit  example for the  problem  of a charge in the constant magnetic field is discussed in detail, and the exact agreement with the Schwinger relativistic form is demonstrated.  A similar analysis is performed in the relativistic Coulomb problem, supporting the accuracy of the proposed relativistic formalism.  
 \end{abstract}

 \section{Introduction  } 
 
 The problem of bound state wave functions and eigenvalues is central in the hadron physics and in QCD in general, and it is especially complicated due to the relativistic character of all interactions, provided by the field  theory.

Historically the  interaction generated by particle exchanges was considered as a basic and the corresponding fundamental equation, the Bethe-Salpeter equation \cite{1,2,3} is used till now  with the kernels, consisting of diagrams with combinations  of particle exchanges, described by the  corresponding propagators. In this way  the Bethe-Salpeter equation corresponds to the summation of an infinite set  of  graphs  of the ladder type, for a review   see  \cite{4*}. 
 
 In the special case, when these propagators are treated instantaneously, e.g. for the virtual photon exchanges, one can use the instantaneous approximation  for the  relativistic Coulomb problem. A  specific feature of the Salpeter equation \cite{2,3} is that it    can be  made free of the aggravating relative time (or zero component of relative momentum) problem.
  
  One special feature of QCD is that the basic  interactions, e.g. confinement, are given  by the vacuum  background fields, and the  latter are well described by the  instantaneous interaction terms, generated by the average Wilson loops, to be called  $ V_{\rm inst}$ (see Appendix for the definition of $V_{\rm inst}$). In  addition there are  other interaction terms,corresponding to particle exchanges, which may be  denoted as operators $\hat V_{op}$, and  actually  contributing  additional terms of the wave  functions, known as Fock column.

 At this point one should stress the difference in the approach of the present paper and numerous   investigations of the relativistic bound state problem in the literature \cite{5*,6*,7*,8*,9*,10*,11*,12*,13*,14*,15*},  see   \cite{4} for a review.

 In  the standard approach the strong interaction is  introduced via the Bethe-Salpeter (Salpeter) equation, or  via Schwinger-Dyson equation,  and one postulates the form of the interaction kernel, responsible for confinement,  trying to find the spectrum by solving integral equations in the  momentum space. 
 
  In a more practical approach (see e.g. \cite{9*}) one postulates the  square root Hamiltonian with confinement and color Coulomb interaction and  finds the hadron spectrum using several adjustable parameters.
  
  In our approach one starts with the basic properties of the field theory in the nonperturbative vacuum.

 In QCD the  confinement is the central problem  which generates masses of  the 99\% of the  visible mass  in the Universe, and it is  very important to take it into account from  the very beginning. As it is known  (see \cite{17*} and  a recent review \cite{5}), confinement is brought up  by vacuum field configurations (it is  not important  here  whether by field  correlators or vacuum  magnetic monopole-like configurations) in the Wilson  loops and therefore confinement can be  described   by the instantaneous interactions in these loops.
 
 It is important, that from the Wilson loops and  vacuum  background one obtains both confining linear  potential and also color Coulomb interaction together  with  spin-dependent terms, so that the  instantaneous  part is the dominant one, and all other exchanges  one can consider as a correction. Moreover, additional gluon exchanges should be treated as perturbative  terms and not  instantaneous, since they provide corrections with an account of additional  gluons in the total wave function, and hence  additional Fock components. In this way one can  avoid  the problem of double counting. 
 
 Another  condition arises due to the fact, that the linearly rising confinement cannot be conveniently used in the momentum space, hence one needs relativistic bound state equations with the  instantaneous interaction in the coordinate space, which is our  main goal below.
 
 The formalism  discussed below was introduced 30 years ago \cite{10,10a,10b,10c,10d} and  was actively developed since then, however the rigorous derivation and comparison with  exact examples were missing. The present paper is aimed to repair these defects. 
 
  The paper is organised as follows. In the next section we derive the proper time representation of the  Green's function and the corresponding  Hamiltonian form, where we show that for the time-independent interaction $V_{inst}$ one obtains  a simple  expression for the Green's function,  which is expressed via the  3d nonrelativistic-like form of Hamiltonian and via the corresponding Green's function,  containing  an integrable parameter.
  
  In section 3 we are  checking this representation in the problem of relativistic Green's function of a charge in  the constant magnetic field. and show that the Schwinger expression for the Green's function \cite{6} exactly coincides with ours. 
 
 In section 4  we are discussing the relativistic Coulomb  problem  and in  section 5 the case of the relativistic harmonic oscillator and  demonstrate the applicability of our approach. In section 6 we generalize the one-body  problem in the  external field to the two-body problem with instantaneous interaction, and obtain the equivalent relativistic Green's functions. We consider   the problem of eigenfunctios and eigenvalues for our Green's functions, which have the  form of  one-dimensional  integrals of  quantum mechanical nonrelativistic  Green's functions. It appears, that the resulting eigenvalues may  approximately coincide with those of the ``square root''   (``spinless Salpeter  equation'') \cite{9*} or  the so-called einbein form of equation \cite{ 10,10a,10b, 10c,  11}.   The character and accuracy of these approximations are discussed in detail. In the final section the main results of the paper are summarized  and the possible areas of applications are discussed. 
 
 \section{Formulation  of the Green's function  representation and the  proper time  Hamiltonian}
 
 We start with the  scalar  particle Green's function in the external field $A_\mu(z)$ and  we write the corresponding proper time  path-integral  representation in the Euclidean space-time \cite{ 10,9,12}
 \be g(x,y) = \left( \frac{1}{m^2-D^2_\mu}\right)_{xy} = \int^\infty_0 ds e^{s(m^2-D_\mu^2)}=\int^\infty_0 ds (D^4 z)_{xy} e^{-K} W (x,y), ~~ D_\mu = \partial_\mu - ig A_\mu,\label{1}\ee
 where 
 \be (D^4z)_{xy} =\int \frac{d^4p}{(2\pi)^4} \prod_k\frac{d^4\Delta z(k)}{(4\pi\varepsilon)^2} \exp \left[ip\left(\sum_k \Delta z (k) - (x-y) \right)\right], ~~N\varepsilon =s\label{2} \ee
 \be K=m^2s + \frac14 \int^s_0 \left( \frac{dz_\mu}{d\tau}\right)^2 d\tau\label{3}\ee
 and the Wilson line $W(x,y)$ is 
 \be W(x,y) = P_A \exp  \left(ig \int^x_y A_\mu dz_\mu\right).\label{4}\ee
 
 In what follows we shall consider the basic type of  interactions, contained in $W(x,y)$, where there is no dependence on time, except that $W $ depends on the points $z$ of the  trajectories $z(t_E)$, i.e. in (\ref{4}) the fields $A_4 (\vez), \veA(\vez)$ do not depend on time, but only on the 3d coordinates.
 
 This condition is  satisfied by constant Coulomb and constant magnetic fields in QED. In the case of QCD we show in the appendix 1  following \cite{9,12}, that both confinement and color Coulomb forces satisfy these criteria at least to the lowest order in $O(\alpha_s)$.
 
 In this case one can rewrite (\ref{4}) as 
 \be W(x,y) =P_A \exp \left(ig \int^{x_4}_{y_4} A_4 (\vez) dt_E +ig \int \veA (\vez) \frac{ d\vez}{dt_E} dt_E\right)\label{5}\ee
 
 As it is clear in (\ref{1}-\ref{4}), the quantum evolution parameters are $\tau$, and $s$, where $0\leq \tau\leq s$, while $z_4(\tau)$ plays the same role as the spatial coordinates $\vez(\tau)$, unless $W(x,y)$ introduces its own $z_4$ dependence. In our case, when $A_\mu(\vez)$ does depend on $z_4$ only via trajectories $\vez (t_E)$, this does not happen and  we  are left with the only evolution parameter $\tau, s$, that one can   connect to the Euclidean time parameters $t_E, T=x_4-y_4$ in the following way 
 \be d\tau = \frac{ dt_E}{2\omega}, ~  s = \frac{T}{2\omega} , ~~ T= x_4-y_4, \label{6}\ee
 and $g(x,y)$ acquires the form
 \be g(x,y) = \sqrt{\frac{T}{8\pi}} \int\frac{ d\omega}{\omega^{3/2}}(D^3z)_{\vex\vey} e^{-K(\omega)} W(x,y), \label{7}\ee
 
 \be K(\omega)= \int^T_0 dt_E \left( \frac{\omega}{2} + \frac{m^2}{2\omega} + \frac{\omega}{2} \left( \frac{d\vez}{dt_e}\right)^2\right).\label{8}\ee
 
 As a next step one can introduce in the path integral (\ref{7}) the Hamiltonian \cite{12}, obtaining in this way for $W\equiv 1$
 \be  \int (D^3 z)_{\vex\vey} e^{-K(\omega)} = \lan \vex | e^{-H_0 (\omega)T}|_{\vey} \ran \equiv f_0 (\vex, \vey)\label{9}\ee
 with \be H_0(\omega) = \frac{\vep^2 } {2\omega}+ \frac{m^2+\omega^2}{2\omega}, ~f_0 (\vex,\vey) = \left( \frac{\omega}{2\pi T} \right)^{3/2}\exp \left( -\frac{m^2T}{2\omega} - \frac{(\vex-\vey)^2}{2T} \omega - \frac{\omega t}{2} \right).\label{10}\ee
 
 In the general case with the account of $W$ one has 
 \be g(x,y) = \sqrt{\frac{T}{8\pi}} \int^\infty_0 \frac{d\omega}{\omega^{3/2}} \lan \vex |e^{-H (\omega) T}|_{\vey}\ran \label{11}\ee
 where $H(\omega)$ using (\ref{5}) can be written as 
 \be H(\omega) = \frac{(\vep-e\veA)^2}{2\omega} + \frac{m^2+\omega^2}{2\omega} +eA_0\label{12}\ee
Using (\ref{1}) and (\ref{9}) one can write the free Green's function as 
$$ g_0 (x,y) = \sqrt{ \frac{T}{8\pi}} \int \frac{d\omega}{\omega^{3/2}} e^{-\frac{m^2+\omega^2}{2\omega} T} \lan \vex |e^{-\frac{ \vep^2}{ 2\omega} T}|\vey\ran =$$\be \sqrt{ \frac{T}{8\pi}} \int \frac{d\omega}{\omega^{3/2}} e^{-\frac{m^2+\omega^2}{2\omega} T}  g^{(0)}_{nr} (\vex, \vey, T),\label{13a}\ee
where  $  g^{(0)}_{nr}$ is  the nonrelativistic free Green's function for the mass  equal to $\omega$,  \be  g^{(0)}_{nr}= \lan \vex |e^{-\frac{ \vep^2}{ 2\omega} T}|\vey\ran= \left( \frac{ 2\pi \omega}{T} \right)^{3/2} \exp \left( -\frac{(\vex-\vey)^2\omega}{2T}\right).\label{14a}\ee

 In the general case of an arbitrary static interaction 
   one can take into account, that $\lan \vex |e^{-HT}| {\vey}\ran$ is simply the 3d nonrelativistic Green's function $G_{nr} (\vex, \vey, T)$ for the given nonrelativistic Hamiltonian $H(\vep, A_\mu)$, augmented by the term $  \frac{m^2+\omega^2}{2\omega}$, so that we  obtain a remarkable expression for the relativistic Green's function of a particle in time-independent fields
 \be g(x,y) = \sqrt{\frac{T}{8\pi} } \int^\infty_0 \frac{d\omega}{\omega^{3/2}} G_{nr} (\vex, \vey, T).\label{13}\ee
  We shall call this form the $\omega$-representation  (of a Green's function)\footnote{Note, that in the early papers \cite{ 10,10a,10b} the corresponding parameter was named $\mu$.}. 
 In   the next section we shall check this relation in the example of  a particle in constant magnetic field. Here the relativistic $g(x,y)$ is given by Schwinger \cite{6}, while $G_{nr} (\vex, \vey, T)$ is known from textbooks on quantum mechanics \cite{13}. 
 
In the QCD the situation with quark and gluon Green's functions is more complicated  due to confinement, which  requires, that all physical objects are gauge invariant and in SU(N) they obtain  their masses due to  confinement, while  in QCD also  quark masses contribute to the final result.

The most important requirement of confinement is that any Green's function of colored   (nongauge invariant) system vanishes after vacuum averaging.

Therefore one cannot define the physical (i.e. gauge invariant and measurable) Green's function for a gluon or a quark.

Similarly any Hamiltonian (or any operator with measurable spectrum) in QCD must be  gauge invariant, and e.g. the  often used operator $(m^2+\hat D^2)$ is not gauge invariant and does not have physically  measurable spectrum in the confining QCD vacuum. Therefore the gauge invariant Green's functions with physical spectrum can be formed using gauge invariant combinations, like $\lan \bar q \Gamma q (x) \bar q \Gamma q (y) \ran $,  $\lan tr ( F_{\mu\nu} F_{\lambda \sigma} (x)) tr F_{\mu x} F_{\lambda \sigma} (y) \ran $ etc., or taking  one of $q$ with infinite mass, one has gauge invariant Green's function of heavy-light mesons $G_{hl} (x,y) = \lan \bar q (x) \Phi (x,y) q(y) \ran$. For gluons one may have a more complicated gauge invariant  structure for the gluelump Green's functions  $G_{gl} (x,y) = \lan tr F_{\mu\nu} (x) \Phi^{adj} (x,y) F_{\lambda\sigma} (y) \ran$, where $\Phi(x,y)$ is the same as $W$, Eq. (\ref{4})  for fundamental representation. In what follows we shall consider for simplicity first the case of  the  heavy-light meson, so  that the local potentials $V_i(r)$ depend on the distance $r$ between quark and the infinitely massive static antiquark.

 We now turn to  the case of QCD  where the main  interaction is due to vacuum fields,  which produce both confining potential $V_{\rm conf}(\ver)$ and  the color Coulomb interaction $V_{\rm coul} (\ver)$, as shown in appendix 1. In this case  we consider a heavy-light system, where neglecting spin forces one can write the Hamiltonian $H(\omega)$ as 
 \be H_{HL} (\omega) = \frac{\vep^2+ m^2 + \omega^2}{2\omega} + V_{\rm conf} (\ver) + V_{\rm coul}(\ver)\label{14}\ee
 with \be V_{\rm conf} (\ver) = \sigma r, ~~ V_{\rm coul}(\ver) = - \frac43 \frac{\alpha_s (r)}{r}.\label{15}\ee

 At this point one may have two different  strategies  to find relativistic eigenvalues and eigenfunctions from (\ref{13}),(\ref{14})  using the corresponding nonrelativistic functions. In the first approach using  the  expansion
 
 \be G_{nr} (\vex, \vey, \omega, T)= \sum_{n} \varphi_n^{(nr)} (\vex, \omega ) e^{-\varepsilon^{(nr)}_n (\omega)  T} \varphi_n^{(nr)} (\vey, \omega)\label{16}\ee
one has to integrate over $d\omega$ in (\ref{13}) using the stationary point method, which yields
\be g_{\rm rel} (x,y) \sim \sum \varphi_n^{(nr)} (\vex, \omega_0 ) e^{-E_n (\omega_0)  T} \varphi_n^{(nr)} (\vey, \omega_0) \label{17}\ee where
\be E_n (\omega_0, T) = \frac{m^2+\omega^2_0}{2\omega_0} + \varepsilon_n^{(nr)}(\omega_0)\label{18}\ee
and $\omega_0$ is to be found from the condition $\left.\frac{\partial}{\partial\omega}(E_n (\omega))\right|_{\omega=\omega_0} =0$.
 This type of approach was named ``the einbein method'' \cite{ 10,10a,10b,10d,11}, and the corresponding eigenfunctions  and eigenvalues were widely used in the hadron physics 
 
 Another approach exploits the form 
 \be G_{nr} (x,y, T,\omega) = \lan \vex |e^{-H(\omega) T} |\vey\ran\label{19}
 \ee
 and suggests to find the stationary point of the operator
 
 \be { H} (\omega, T) = \frac{m^2+\omega^2}{2\omega} +\frac{\vep^2}{2\omega} + V(r).\label{20}\ee
 In the case when the instsntaneous  interaction $V(r)$ does not depend on $\omega$, one obtains a simple answer, usually called the spinless  Salpeter  or  the square root kernel 
 \be { H} (\omega_), T) = \sqrt{\vep^2+m^2}+ V(r),\label{21}\ee which is commonly used in relativized quark models \
 \cite{5*,6*,7*,8*,9*,10*}, and is important in the dynamics of Regge trajectories \cite{25*,25a,25b,25c,25d,25e}.
  
 As we shall see below, in general the ``einbein'' eigenvalues (\ref{18}) and the ``spinless Salpeter type'' eigenvalues of (\ref{21}) may not coincide and moreover the operator (\ref{21}) may need renormalization \cite{16,17}, in the case when $V(r)$ is singular and may produce singular eigenfunctions. This happens e.g. in the case of Coulomb interaction which we shall discuss below in section 4.  
 
 \section{Relativistic Green's function in a constant magnetic field}
 We start with the relativistic Green's function of a scalar particle in a constant magnetic field, which can be taken from Schwinger \cite{6} in our notations as
 \be G_S(x,y) =\frac{1}{(4\pi)^2} \int^\infty_0 \frac{ds}{s^2} e^{-im^2s} \Phi(x,y) e^{-L(s)} \exp \left[ \frac{i}{4} (x-y)_\mu C_{\mu\nu} (x-y)_\nu\right]\label{3.1}\ee
 where $L(s) =\ln \left(\frac{\sin eBs}{eBs}\right), $ 
 $$ C_{\mu\nu} = \frac{1}{s} (\delta_{\mu 3 }\delta_{\nu 3} - \delta_{\mu 0}\delta_{\nu 0}) + eB ~  {\rm ctg} ~(eBs )( \delta_{\mu 1}\delta_{\nu 1}   + \delta_{\mu 2}\delta_{\nu 2});$$~ $\Phi(x,y) = \exp\left( ie\int^x_y du_\mu A_\mu(u)\right)$.
 
 The nonrelativistic Green's function of the scalar particle with mass $\omega$ in the magnetic field $B$ along the 3d axis is 
 $$ G_{NR} (x,y, \omega) = \left(\frac{\omega} {2\pi it} \right)^{3/2}  \left(\frac{\frac{eBt}{2\omega}}{\sin \frac{eBt}{2\omega}}\right) \exp \frac{i\omega}{2t} (x_3-y_3)^2\times$$\be \times \exp \left\{\frac{ieB }{4} \left[ {\rm ctg}~ \frac{eBt}{2\omega}  \left((x_1-y_1)^2+(x_2-y_2)^2\right)\right]^4 \right\}\label{3.2}\ee
 where we have chosen the phase $\Phi\equiv 1$. One can check, that for $eB \to 0$ one obtains the free nonrelativistic Green's function. 
 Writing now the relativistic Green's function as in (\ref{11})  for $it =T$ in the Euclidean space-time,
 \be G_{\rm Rel} (x,y,T) = \sqrt{ \frac{T}{8\pi}} \int^\infty_0\frac{d\omega}{\omega^{3/2}} e^{-\frac{m^2+\omega^2}{2\omega} T} G_{NR} (x,y,T, \omega) \label{3.3}\ee
 to reproduce the Schwinger result (\ref{3.1}) one should write in (\ref{3.2}) $\frac{\omega}{it} = \frac{\omega}{T} = \frac{1}{2s}$,    and one obtain the identity of (\ref{3.3}) and (\ref{3.1}).
 
 We do not consider here the spin degrees of freedom, indeed, for the bispinor particle the resulting expression cannot be simply obtained in the form of (\ref{11}), since these degrees of freedom are  not present in the nonrelativistic Green's function. However the simple magnetic dipole interaction, which appears for  electron in the Schwinger expression as an extra factor $\exp \left[ i \frac{e}{2} \sigma_{\mu\nu} F_{\mu\nu} s\right]$ can be  easily reproduced if in the nonrelativistic Hamiltonian $H(\omega)$ (\ref{12}) one adds  a similar term $\Delta H(\omega) = \frac{e\sigma_{\mu\nu} F_{\mu\nu}}{2\omega} T$  to account for the spin interaction.
 
 Finally, in QCD the numerator of the spinor or vector Green's function is treated separately as shown in \cite{12} and  again is expressed via $\omega$ and field correlators, and the resulting expressions e.g. for meson coupling constants, e.g. $f_\rho, f_D, f_{D_s}, f_B$ etc. are in very good  agreement with lattice data and experiment \cite{18}. 
 
 We now come to the problem of eigenvalues.  To this end we calculate the momentum transform of the Green's function (\ref{3.1}) and (\ref{3.3}).  From (\ref{3.1}) one obtains \cite{37*}
 \be G_S (k) = \int^\infty_0 ds \exp \left[ -ism^2 +is (k^2_0-k^2_z) - is \vek^2_\bot \frac{{\rm tang} (eBs)}{eBs} \right] = \int d^4 (x-y)e^{ik(x-y)} G_S (x-y). \label{3.4}\ee

 This can be expanded as a sum over Landau levels
 \be G_S(k) =- \sum^\infty_{n=0} \frac{2(-1)^n e^{-\alpha_k} L_n (2\alpha_k)}{k^2_0-k^2_z-m^2- (2n+1) eB+i\varepsilon},\label{3.5}\ee
 where $\alpha_k = - \frac{k^2_\bot}{eB}$, and $L_n(x)$ is the Laguerre polynomial. 
  
 Another, and  a more simple way  to find eigenvalues of the relativistic Green's function is  to write $G_{NR}$ in  (\ref{3.3}) via  the Hamiltonian in magnetic field, as in (\ref{11}), (\ref{12})
 
 \be 
 G_{NR} (x,y,T, \omega) =\lan\vex |e^{-H_{\rm mag} (\omega) T}|_{\vey}\ran \label{3.6}\ee
 where  $H_{mag}$ using (\ref{12}) can be written  in a standard way
 \be H_{\rm mag} (\omega) = \frac{\vep^2}{2\omega} + \frac{e^2 (\veB\times \ver)^2}{8\omega} - \frac{e\vesig \veB}{2\omega }+ \frac{ m^2+\omega^2}{2\omega}, \label{3.7}\ee
 where we have added the spin interaction term to apply our result also to a fermion.
 
 One can immediately write the lowest eigenvalues of the Hamiltonian (\ref{3.7}), known from nonrelativistic quantum mechanics \cite{13}
 \be \varepsilon^{\rm non}_n (\omega) = \frac{p^2_z + eB (2 n +1) - e\vesig\veB}{2\omega}. \label{3.8}\ee
 Using (\ref{3.3}) one can write $G_{Rel} (x,y,T)$ as follows
 \be G_{\rm Rel} (x,y,T) =\sqrt{ \frac{T}{8\pi}} \int^\infty_0\frac{d\omega}{\omega^{3/2}} e^{-\frac{m^2+\omega^2}{2\omega} T} \sum_n \varphi_n (\omega, x) e^{-\varepsilon^{\rm nonr}(\omega ) T} \varphi^+_n (\omega, y).\label{3.9}\ee

 The resulting integral (\ref{3.9}) can be estimated at large $T$ using the stationary point method, which yields
 \be \left.\frac{\partial}{\partial\omega} \left( \frac{m^2+\omega^2+\varepsilon^{\rm nonr}_n (\omega)}{2\omega}\right)\right|_{\omega=\omega_0}= 0,~~ \omega_0 = \sqrt{m^2+p^2_z + eB(2n+1)-e\vesig \veB}\label{3.10}\ee
 and the einbein  eigenvalues are 
 \be E_n = \left.\left( \frac{m^2+\omega^2+\varepsilon^{\rm nonr}_n (\omega)}{2\omega}\right)\right|_{\omega=\omega_0}=\sqrt{m^2+p^2_z + eB(2n+1)-e\vesig \veB}.\label{3.11}\ee
 This result exactly coincides with the eigenvalues of the relativistic Green's function (\ref{3.5}) for scalar particle, when $e\vesig \veB$ is absent, and with the corresponding eigenvalues of a fermion with account of $\vesig \veB$.

 Note, that the same eigenvalues obtain in the  square-root (spinless Salpeter) method, when one first find the stationary point Hamiltonian $H_{\rm mag}(\omega_0)$ from (\ref{3.7}).
 \section{The case of the (color) Coulomb interaction}
 
 We consider a scalar particle in the Coulomb field $ V_c(r) = - \frac{Z \alpha}{r}$, and using the Klein--Gordon equation following \cite{19}, one has 
 \be \left[ \left( E + \frac{Z\alpha}{r} \right)^2 + \Delta - m^2 \right) \phi =0\label{4.1}\ee
 which can be rewritten as 
 \be \left\{ - \frac{\partial^2}{\partial r^2} -\frac{2\partial }{r\partial r } + \frac{L^2 -Z^2\alpha^2}{r^2} - \frac{2Z\alpha E}{r} -(E^2-m^2)\right\} \phi(r) =0.\label{4.2}\ee 
 
 The  resulting energy eigenvalues can be written as 
 \be E_{nl} = \frac{m}{\sqrt{1+ \frac{Z^2\alpha^2}{(n-\delta_l)^2}}},\label{4.3}\ee
 here $\delta_l$ is 
 \be \delta_l = l+1/2-\sqrt{(l+1/2)^2 - Z^2\alpha^2}.\label{4.4}\ee
 
 On the other hand, one can use as in (\ref{3.2}) the nonrelativistic Hamiltonian
 \be H=\frac{\vep^2}{2\omega} + V_c (r)+ \frac{m^2+ \omega^2}{2\omega} \label{4.5}\ee
which yields eigenvalues
\be \varepsilon_n(\omega) = - \frac{\omega(Z\alpha)^2}{2n^2}\label{4.6}\ee
and one can find the extremum  in $\omega$ of the total eigenvalue 
\be M_n (\omega) = \frac{m^2+\omega^2}{2\omega} - \frac{\omega(Z\alpha)^2}{2n^2}\label{4.7}\ee
yielding 
\be M_n (\omega_0) = m\sqrt{1-\left( \frac{Z\alpha}{n}\right)^2}\label{4.8}\ee
which differs from (\ref{4.3}), since we have treated in (\ref{4.5}) the Coulomb field as a scalar, while in (\ref{4.1}) its vector character is accounted for.

To take into account the vector character of  the Coulomb interaction we shall use the ``square root approach'' to the  full Hamiltonian (\ref{4.5})
\be \left. \frac{\partial  H}{\partial \omega} \right|_{\omega=\omega_0} =0, ~~ \omega_0 =\sqrt{\vep^2+m^2}\label{4.9}\ee
which yields the square root Hamiltonian
\be H(\omega_0) =\sqrt{\vep^2+m^2} - \frac{Z\alpha}{r}; ~~ H(\omega_0) \Psi_n = \tilde{E}_n  \Psi_n\label{4.10}\ee
and the resulting equation for $\tilde{E}_n$ has the same form as in (\ref{4.1}), when one takes the quadratic expression
\be \left( \sqrt{\vep^2 + m^2}\right)^2\Psi_n = \left( E_n + \frac{Z\alpha}{r}\right)^2 \Psi_n\label{4.11}\ee
with  the same spectrum (\ref{4.3}).

Turning now to the case of the relativistic fermion in the electromagnetic field, our hamiltonian $H(\omega)$ has the form \cite{16}
\be H(\omega) =\frac{(\vep-e\veA)^2}{2\omega} + \frac{m^2+\omega^2}{2\omega} + e A_0 - \frac{e \vesig \veB}{2\omega} - \frac{ie(\veal \veE)}{2\omega}\label{4.12}\ee
which simplifies when $\veA = \veB =0$. Here $\veal = \left( \begin{array}{ll} 0&\vesig\\\vesig&0\end{array} \right)$;
\be H_{ll} (\omega) = \frac{\vep^2}{2\omega} + \frac{m^2+\omega^2}{2\omega} - \frac{Z\alpha}{r} - ie\frac{(\veal \veE)}{2\omega}. \label{4.13}\ee
Following the same way as in (\ref{4.9})-(\ref{4.11}), but now with the additional last term in (\ref{4.13}), one obtains the exact Dirac spectrum, as it was shown in \cite{16}, 
\be M_n = \frac{m}{\sqrt{1+\left( \frac{Z\alpha}{n-\delta_j}\right)^2}},  ~~ n=1,2,...\label{4.14}\ee
where $\delta_j$ is the same as in (\ref{4.4}) with the replacement $l\to j$. Note, that both forms (\ref{4.8}) and (\ref{4.14}) coincide  to the lowest order $O\left((Z\alpha)^2\right)$. 

\section{ The case of the relativistic harmonic oscillator interaction}

Here we consider the scalar particle with the harmonic oscillator interaction of the vector or scalar type. In the first case one has the Klein-Gordon equations similar to (\ref{4.1}), but with the replacement $-\frac{Z\alpha}{r} \to c r^2, ~~
c>0$, which yields
\be   (\vep^2 + m^2- (c r^2)^2 + 2 E cr^2)\varphi =  E^2\varphi.\label{5.1}\ee

This equation evidently has no lower bound for the eigenvalues and hence no discrete spectrum.

This conclusion 
agrees with the case of the Dirac equation with the harmonic oscillator interaction of the  vector type, considered in \cite{20}, where it was shown both numerically and analytically that no discrete spectrum appears in this case.

Let us turn now to the case of the scalar harmonic oscillator interaction, in which case the wave equation has the form
\be [E^2-\vep^2- (m^2+cr^2)]\varphi =0\label{5.2}\ee
with the stable discrete  spectrum
\be E_n = \sqrt{m^2+\sqrt{4c} (n+3/2)}.\label{5.3}\ee
This result again agrees qualitatively with the corresponding Dirac spectrum found in \cite{20}.

Another useful comparison can be made with the case of the scalar particle in the constant magnetic field, where the interaction term is also of the harmonic oscillator type; neglecting the spin term in (\ref{3.11})  one obtains \be E_n({\rm mag}) = \sqrt{ m^2 +p^2_3+ eB (2n+1)}.\label{5.4}\ee

One can see the same type of the spectrum in 2d, while in (\ref{5.4}) additional free  continuous spectrum is present in the third dimension. 

\section{ The quark-antiquark  relativistic Green's functions and energy eigenvalues}

For the $\bar q q$ system the (relativistic) Green's function is obtained as a product of $q$ and $\bar q$ Green's functions with the corresponding vertices,
\be G_{q\bar q} (x,y) = \left\lan \frac{(m_1 - \hat D_1) \Gamma (m_2-\hat D_2) \Gamma}{(m^2_1- \hat D^2_1)( m^2_2- \hat D^2_2)} \right\ran\label{6.1}\ee
and we note, that for the scalar quarks the Green's functions are the denominators in ( \ref{6.1}) without spin operator, i.e. using (\ref{11}) one has  the $\omega$- representation for the $q\bar q$ Green's function (scalar  quarks)
\be \left\lan \frac{1}{ (m^2_1-  D^2_\mu)(m^2_2-  \bar D^2_\mu)}\right\ran_{xy}= \frac{T}{8\pi} \int^\infty_0 \frac{d\omega_1}{\omega_1^{3/2}}\int^\infty_0 \frac{d\omega_2}{\omega_2^{3/2}}\lan \vex, \vex \left| e^{-H(\omega_1, \omega_2, \vep_1, \vep_2)T}\right| {\vey, \vey}\ran\label{6.2}\ee
where 
\be H(\omega_1, \omega_2, \vep_1, \vep_2)= \frac{ \vep^2_1 + m^2_1 + \omega^2_1}{2\omega_1} + \frac{ \vep^2_2 + m^2_2  + \omega^2_2}{2\omega_2}+V(\ver_1\ver_2).\label{6.3}\ee

Now integrating over $d^3(\vex-\vey)$ one obtains the Green's function and the hamiltonian in the c.m. system $\veP= \vep_1+\vep_2=0$, while taking the trace in the numerator of (\ref{6.1}) one obtains finally the  $\omega$-representation for the $q\bar q$ c.m. Green's function \cite{12}

\be G_{\bar q q}^{C.M.} (x,y) = \int G_{\bar q q} (x,y) d^3(\vex-\vey) = \frac{T}{2\pi} \int^\infty_0\frac{d\omega_1}{\omega_1^{3/2}}\int^\infty_0 \frac{d\omega_2}{\omega_2^{3/2}}\lan Y_\Gamma\ran \lan 0   \left| e^{-H_{C.M.}(\omega_1, \omega_2, \vep )T}\right|0\ran\label{6.4}\ee
where \be \lan Y_\Gamma\ran= \frac14 tr \lan \Gamma(m_1-   \hat D_1)  (\Gamma(m_2-   \hat D_2) \ran.\label{6.5}\ee

The spin d.o.f. enter in two ways: 1) in the numerator $\lan Y_\Gamma\ran$ explicitly; 2) via the denominator, where spin-vacuum interaction $ \lan \sigma F\sigma  F\ran $ ensures a strong self-energy correction 
\be (m^2-(\hat D_\mu\gamma_\mu)^2)^{-1} = (m^2- D^2_\mu - g \sigma_{\mu\nu} F_{\mu\nu} )^{-1} = G_0 (1+ G_0 g\sigma F G_0 g\sigma F+...).\label{6.6}\ee

The averaging over vacuum fields in  the last term in (\ref{6.6}) yields the self-energy correction to the Hamiltonian \cite{21}
\be H_{C.M.} \to  H_{C.M.} + \Delta M_{SE}, ~~ \Delta M_{SE} =- \frac{b\sigma\eta_1}{ \pi\omega_1}- \frac{b\sigma\eta_2}{ \pi\omega_2}.\label{6.7}\ee
Here the factor ``$b$'' contains a renormalized contribution of field correlators, we take it  with corrections found in \cite{23},  as shown in appendix 2,    and $\eta_i \equiv \eta(m_i)$ are calculated in \cite{21},  normalized as $\eta_i(0)=1$.
  Note, that the effect of the (negative) $\Delta M_{SE}$ term is very important for the spectrum and  as will be seen below, strongly decreases the mass. A  similar role  in  the  standard Salpeter approach (see e.g. \cite{9*}) is played by   an adjustable negative   constant in the Hamiltonian, which however  was found different for light  and heavy  flavors and moreover violates the linear Regge trajectories. In our case the constant $b$ is defined by field correlators $D^E, D^E_1$ and  does not depend on  the quark masses, while the function $\eta(m_i)$ is well defined for all values of quark masses, see Appendix 2 for details.
  
As  a result the total Hamiltonian with the color Coulomb, self-energy and spin-spin interaction acquires the form \cite{22, 23}, where the spin-spin term was obtained in the framework of the Field Correlator Method in \cite{34}. 

\be 
H=H_0 + V_{\rm conf}+ V_{OGE} + \Delta M_{SE} + \Delta M_{ss},\label{6.8}\ee
where 
\be H_0 = \sum_{i=1,2} \frac{m^2_1+\omega^2_i}{2\omega_i} + \frac{\vep^2}{2\tilde \omega}, ~~ \tilde \omega= \frac{\omega_1 \omega_2}{\omega_1 +\omega_2}\label{6.9}\ee

\be V_{\rm conf} = \sigma r, ~~ V_{OGE} = - \frac{4 \alpha_s (r)}{3r},
~~ \Delta M_{SE} = - \sum_i\frac{ b\sigma \eta ( \omega_i)}{ \pi
\omega_i}\label{6.10}\ee

\be \Delta M_{ss} = \frac{\vesig_1 \vesig_2}{12 \omega_1 \omega_2}
V_4 (r) , ~~ V_4 (r) =\frac{32 \pi \alpha_s}{3} \delta^{(3)}
(\ver).\label{6.11}\ee

We start with the case of one light quark and a heavy antiquark. Neglecting the
contribution of the   latter, one obtains the one-particle problem in  the
linear potential, which we shall treat in  our formalism and compare the
spectrum to the  Dirac equation with linear potential, and at the end discuss
its relation to the heavy-light meson. In our present formalism we have two
possibilities  for the Hamiltonian.

 \begin{enumerate}
\item 
The so-called einbein form \cite{10,12}

\be \left( \frac{m_i^2+\omega^2_1}{2\omega_1} +
\frac{\vep^2}{2\omega_1} + \sigma  r +{\Delta M_{SE}}\right)
\psi_n = M_n \psi_n.\label{6.12}\ee
\item
The square-root form (the spinless Salpeter equation)  

 \be \left(\sqrt{\vep^2 + m^2_1} + \sigma r +\Delta
M_{SE}\right)\psi_n = M'_n \psi_n.\label{6.13}\ee
\end{enumerate}
 In the case 1, taking into account, that eigenvalues $\varepsilon_n$ of the
equation \be \left( \frac{\vep^2}{2\omega} + \sigma r\right)
\varphi_n = \varepsilon_n \varphi_n\label{6.14}\ee are equal to
$\varepsilon_n = \frac{\sigma^{2/3} a(n)}{(2\omega)^{1/3}} , ~~ a
(0) = 2.338$,  one obtains for $m=0$ \be M_0 = 0.78~ {\rm GeV},~~
M'_0=0.72 ~{\rm GeV} ~(b=2).\label{6.15}\ee

It is interesting to compare these values with the result for the ground state
level of the Dirac equation with linear potential,  which was done in \cite{20},

\be M_0^D = 0.689 ~ {\rm GeV}\label{6.16}\ee

One can see a  a good agreement within 30 MeV of the spinless Salpeter result $M'_0$ with the Dirac eigenvalues (\ref{6.16}).

One can realize,
that both equations (\ref{6.12}),(\ref{6.13}) contain only positive energy levels, and
no contribution from negative energy levels is present in $M_n,
M'_n$, whereas in $M^D_n$ all 4 components of the wave function
contribute.

Therefore in  heavy-light case one can see an additional (Dirac) eigenvalue suppression as compared to the standard self-energy mechanism.

We now  turn back to the case of the $q\bar q$ system in the confining vacuum. We again consider the Hamiltonian (\ref{6.8}) with the $\Delta M_{SE}, ~ V_{OGE}$ and $\Delta M_{ss}$ terms and take into account the color Coulomb interaction also in calculating $\omega$, which is important for the lowest $q\bar q$ states with small masses $m_q, m_{\bar q}$, e.g. for the $\rho$ meson.

In this case in  the  approach 2) (square root Hamiltonian) one obtains \cite{22}
\be \bar \omega = \omega_{\rm Salp} + \Delta \omega_{\rm coul}= (0.335 +0.054) {\rm ~GeV} = 0.39  {\rm ~GeV}\label{73}\ee
where $ \Delta \omega_{coul}$ is the change  of $\omega$ due to  color Coulomb interaction. 
\be \Delta M_{SE} = - \frac{3\sigma}{\pi\bar \omega} \cong - 0.415 {\rm ~GeV},~~ M_0 =1.339 {\rm ~GeV}\label{74}\ee

\be \lan V_{OGE}\ran = - \frac34 \alpha_s \lan r^{-1}\ran = -0.60 \lan r^{-1}\ran=-0.218 {\rm ~GeV}\label{75}\ee
 and the resulting spin-averaged mass is 
 \be \bar M = M_0 + \Delta M_{SE} + \lan V_{OGE} \ran = 0.706 {\rm ~GeV}.\label{76}\ee
Finally one must take into account the hyperfine interaction, which yields
\be \Delta_{ss} = \frac{32 \ves_1\ves_2 \alpha_s \pi}{9\bar \omega^2} \varphi^2_n (0)\to \frac{8\alpha_s\pi}{9 \bar \omega^2} \varphi^2_n (\omega), \label{77}\ee
where $\varphi^2_n(0)$ can be estimated as follows, 
\be \varphi^2_n (0) = \frac{\bar \omega}{4\pi} \left( \sigma + \frac43 \alpha_s\left \lan \frac{1}{r^2}\right\ran \right) \cong \frac{0.109}{4\pi}{\rm ~GeV}^3\label{78}\ee
 and as a result one obtains for $\rho$ meson
 \be \Delta_{ss} (\rho) = (0.048\div 0.065) {\rm ~GeV},\label{79}\ee
 depending on  whether one is using Coulomb-corrected (left) or noncorrected value of $\omega$ (right).
 
 Finally one obtains
 \be M_{\rho} ({\rm ~ square~ root~ Hamiltonian}) = (0.754-0.771) {\rm ~GeV}\label{80}\ee
 to  compare with the experimental value $M_\rho(\exp) =0.775$ GeV.
 
 In a similar way our $\omega$ technique allows to calculate all light meson masses, using only string tension $\sigma, \alpha_s (\Lambda_{QCD})$ and current quark masses, as shown in \cite{ 10, 10a,10b, 10d,25*,25a,25b,25c,22}, without extra parameters. One should also stress that also orbital and radial linear Regge trajectories are obtained in the same formalism exploiting exact representation of the relativistic  Green's functions, and the corresponding eigenvalues, see \cite{22} for more results and references. 
 
 \section{Conclusion}
 
 We have derived above the exact relativistic Green's functions  for the static (time-independent) interactions, which allow to write down simple integral representation  the $\omega$-representation in terms of the nonrelativistic Green's function and Hamiltonian. These Greens functions and eigenvalues were compared to the known relativistic results of the constant magnetic field and Coulomb interaction also including spin, and the agreement was demonstrated.
 
 In this way we can  exploit the vacuum generated  instantaneous interaction $V_{conf}$ and $V_{coul}$ in the $\omega$-representation to obtain the  basic  results for masses and wave functions,  leaving additional gluon and $q\bar q$ creation terms as a perturbation. 
 
 As a practical implementation the masses of heavy-light and light-light mesons have been considered and a close correspondence with experiment in the case of the $\rho$ meson was shown in detail. It is important,  that this method exploits only few basic parameters: $\sigma, \alpha_s(\Lambda_{QCD})$ and quark current masses, in contrast to other approaches with many additional parameters.
 
 It is possible to make our theory even more fundamental, since  one can  express $\Lambda_{QCD}$ via $\sigma$, as it was shown in \cite{5}, leaving $\sigma$ and current quark masses as the only scale parameters in QCD.

At this point  it is  important to  disclose the meaning of the term ``exact'' in the exact agreement of the  results of our $\omega$- representation and the Schwinger derivation, or  Dirac spectrum. It  is clear, that the fully exact  result e.g. in QCD should contain contribution of additional $q\bar q$ pairs and any number of gluons interacting with initial number of particles, which create the so-called Fock column of wave functions, and this is not provided by the $\omega$-representation or the Schwinger expression for the Green's function, where the electric  charge enters only in the combination $eB$. Therefore one should associate the exact agreement with the coincidence of the corresponding Taylor series in powers of the external field.

 The method, based on the $\omega$-representations discussed above, was developing successfully during the last 30 years and has allowed to calculate analytically mesons, glueballs,  hybrids and baryons in terms of the minimal number of parameters $\sigma, \alpha_s, m_q$), for the latest development see \cite{22}, see also \cite{44,45,46*,47,48} for similar methods and \cite{49,50} for  reviews.

 The author is grateful for help and useful advices to A.M.Badalian. Useful discussions with R.A.Abramchuk, M.A.Andreichikov and Z.V.Khaidukov are gratefully  acknowledged. \\

\setcounter{equation}{0}
\renewcommand{\theequation}{A1.\arabic{equation}}

\noindent
 {\bf \large  Appendix  1  }\\

 The starting point in the vacuum averaged Wilson loop, which we shall take simplicity in the  {\bf 1,4}  plane and keeping the lowest, $\lan FF\ran$ correlator, one has 
 \be \lan W(C)\ran = \exp \left( -\frac{g^2}{2} \int d \sigma_{14} (u_1, u_4) d\sigma_{14} (v_1, v_{14}) \lan F F\ran\right). \label{A.1}\ee
 
 We shall show below, that it can be expressed via instantaneous potentials $ V_D, V_1$. Here $\lan FF\ran$ is 
 
 $$g^2\lan FF\ran =\frac{g^2}{N_c} \lan tr_f (F_{14} (u) \Phi(u,v) F_{14} (v) \Phi(v,u))\ran =$$
 \be = D^E (u-v) + \frac12 \left( \frac{\partial}{\partial u_1} \left[ (u_1-v_1) D^E_1 (u-v) \right] + \frac{\partial}{\partial u_4}  \left[ (u_4-v_4) D^E_1 (u-v) \right]\right). \label{A.2}\ee
 
 Insertion of (\ref{A.2}) in (\ref{A.1}), and integrating 
 \be \int d\sigma_{14} (u_1,u_4) = \int^R_0 du_1 \int^T_0 du_4,\label{A.3}\ee
 one obtains using the geometry in Fig.1 

\begin{center}

\begin{figure}
\begin{center}
  \includegraphics[width=4cm, ]
  {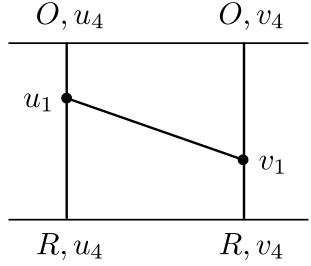}

\end{center}
  \caption{ Calculation of the static potentials $V_D, V_1$ from field correlators}

\end{figure}
 \end{center}
 
  \be \lan W(C)\ran  = \exp      \left (-\int^T_0 \left(V_D (R) + V_1(R)\right)dt \right)\label{A.4}\ee
  where, denoting $w_1=u_1-v_1$, one has 
 \be V_D (R) = 2 \int^R_0 (R-w_1) dw_1 \int^T_{0} dw_4 D^E  \left(\sqrt{w^2_1+ w^2_4}\right) \label{A.5}\ee
 
 \be V_1(R) =  \int^R_0  w_1  dw_1 \int^T_{0}  dw_4 D^E_1  \left(\sqrt{w^2_1+ w^2_4}\right)\label{A.6}\ee
 as it was shown in \cite{10}, $D_1(x)$ at small $x$ has the form 
 \be D_1(x) = \frac{4C_2\alpha_s}{\pi} \left\{ \frac{1}{x^4} + \frac{\pi^2 G_2}{24 N_c} + ... \right\}\label{A.7}\ee
 and therefore it contributes to the color Coulomb interaction, $V_1^{OGE} (r) =- \frac{C_2\alpha_s(r)}{r}$, which is properly renormalized at small distances, see \cite{35*} and \cite{36*} for the analytic and lattice analysis, 
 
 It is important, that actually $V_D, V_1$ are nonlocal both in time and space, and the final local terms (\ref{A.5}), (\ref{A.6}) are obtained in the ``local limit'', when the correlation length in $D^E, D_1^E$ is going to zero, actually $D(w)= f\left( \frac{w}{\lambda}\right), ~~\lambda\sim 0.1$ fm. Hence $V_D, V_1$ are local and instantaneous with this accuracy.\\

\setcounter{equation}{0}
\renewcommand{\theequation}{A2.\arabic{equation}}

\noindent
 {\bf \large  Appendix  2  }\\
 
 Following \cite{21, 40*}, one can use in (\ref{6.6}) the expansion in powers of $(g\sigma_{\mu\nu} F_{\mu\nu})$ and take into account the vacuum  averaging, which keeps nonzero the second order term, which be  represented as 
 \be \Lambda = \int d^4 (u-v) g^2 \lan \sigma F(u) \Phi (u, v)  \sigma F(v)  \Phi (v, u) \ran G(u, v) \label{A2.1}\ee
where $G(u,v) = (m^2_q - D^2_\mu)^{-1}_{u,v}$  is the spinless quark  Green's function.

One can use for the field correlator in (\ref{A2.1})  the  standard representation  (\ref{A.2}) in terms of  correlators $D^E ( u-v)$   and $D^E_1(u-v)$, which have a small correlation length $\lambda \approx 0.2$ fm and decay exponentially at large distances,  $D^E(w) =  D^E (0) \exp (- |w|/\lambda)$. At the same time $G(u,v)$ at small distances $|u-v|\la \lambda$ can be replaced by the corresponding free form 
\be G(u,v) \approx G_0 (u-v) = \frac{m_q K_1 (m_q|w|)}{4\pi^2 |w|}, ~~ w=u-v. \label{A2.2}\ee

As a result $\Lambda$ can be  brought to the form 
\be \Lambda = m_q (3D (0) + 3D_1(0)) \varphi (m_q  \lambda)\label{A2.3}\ee where \be \varphi(m_q\lambda) = \int^\infty_0 w^2 dw e^{-w/\lambda} K_1 (m_q w)
\label{A2.4}\ee and using the relation $\sigma = \frac12 \int  D(x) d^2 x = \frac{\pi D(0)}{\lambda^2}, $ one obtains
\be \Delta M_{SE} =-\frac{\Lambda}{2\omega} =-\frac{3\sigma}{2\pi\omega} \eta (m_q \lambda) (1+ \xi).\label{A2.5}\ee
Here $\eta(m_q \lambda) = \frac{m_q}{\lambda^2} \varphi (m_q \lambda); ~~ \eta (0) =1$, and $\xi = \frac{D^E_1(0)}{D^E(0)}$, denotes the contribution of the correlator $D^E_1(0)$, which is below $\frac13 D^E(0)$,  according to the lattice measurements \cite{46}.

As a result the constant ``b'', introduced in (\ref{6.7}), according to (\ref{A2.5}) is  \be b=\frac32(1+\xi), ~~ 0<\xi \leq 1/3\label{A2.6}\ee
where the parameter $\xi$ is not known with  good accuracy.

The function $\eta (m_q\lambda)$ is normalized as $\eta (0) =1$, and it has asymptotics $\eta \approx \frac{2}{m^2_q\lambda^2}$ at large $m_q$. E.g. for the $b$ quark, $m_q \cong 5$ GeV one has $\eta (m_b \lambda) \cong 0.052$, which makes the contribution of $\Delta M_{SE}$ unimportant, $\Delta M_{SE} (m_b) \approx - 0.89$ MeV.


\end{document}